\documentclass[a4paper,11pt]{scrartcl}\usepackage[centertags]{amsmath}
\usepackage{amsfonts}
\usepackage{amssymb}
\usepackage{multirow}
\usepackage{amsthm}
\usepackage{nicefrac}
\usepackage{pifont}
\usepackage[ansinew]{inputenc}
\usepackage{graphicx}

\begin{document}
\author{Ralph Brinks\footnote{rbrinks@ddz.uni-duesseldorf.de}\\
Institute for Biometry and Epidemiology\\German Diabetes Center\\
Düsseldorf, Germany}
\title{Fast Calculation of Calendar Time-, Age- and Duration Dependent Time at Risk in the Lexis Space}

\date{}

\maketitle

\begin{abstract}
In epidemiology, the person-years method is broadly used to estimate the incidence 
rates of health related events. This needs determination of time at risk stratified by period, age
and sometimes by duration of disease or exposition. The article describes a fast method for 
calculating the time at risk in two- or three-dimensional Lexis diagrams based on Siddon's algorithm. 
\end{abstract}

\emph{Keywords:} person-years method, Lexis diagram, Siddon's algorithm.

\section{Lexis diagram and person-years method}
\begin{figure}[th]
\begin{center}
\includegraphics[width=9.6cm, keepaspectratio]{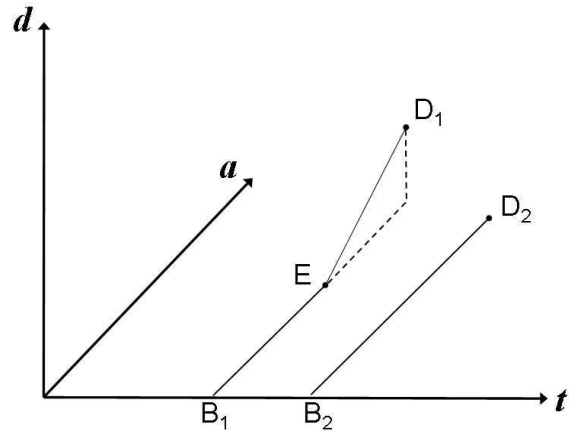}
\end{center}
\caption{Three-dimensional Lexis diagram with two life lines. Abscissa,
ordinate and applicate represent calendar time $t$, age $a$ and duration $d$, respectively.
The life lines start and end at birth $B_n$ and death $D_n, ~n=1,2.$
The first subject gets the disease at $E$. Then, the life line changes its direction.
The second subject does not get the disease, the corresponding life line remains in 
the $t$-$a$-plane.}\label{fig:Lexis}
\end{figure}

In epidemiology, oftentimes relevant events or outcomes simultaneously 
depend on different time scales: age of the subjects, calendar time and 
duration of an irreversible disease. In event history analysis, \cite{Kei06}, 
a useful concept is the \emph{Lexis diagram}, which is
a co-ordinate system with axes calendar time $t$ (abscissa) and age $a$ (ordinate).
The calendar time dimension sometimes is referred to as period. Each subject 
is represented by a line segment from time and age at entry to time and age at exit.
Entry and exit may be birth and death, respectively, or entry and exit in a
epidemiological study or trial. There are excellent and
extensive introductions about the theory of Lexis diagrams 
(see for example \cite{Kei90}, \cite{Kei91}, 
\cite{Car06} and references therein), which allows to be short here.
When it comes to irreversible diseases, the commonly used two-dimensional 
Lexis diagram with axes in time and age direction may be generalized to a 
three-dimensional co-ordinate system with disease duration $d$ represented by the
applicate (z-axis). If a subject does not get the disease 
during life time, the life line remains in the time-age-plane parallel to the
line bisecting abscissa and ordinate. With other words, the life line for
the time without disease points in the $(1,1,0)$ direction (where the
triple $(t,a,d)$ denotes the co-ordinates in time, age and duration direction, respectively). 
However if at a certain point in time $E$ the disease is diagnosed, the life line
changes its direction, henceforth pointing to $(1,1,1)$. The situation is illustrated 
in Figure~\ref{fig:Lexis}. The life lines of two subjects are shown in the
three-dimensional Lexis space. At time of birth (denoted $B_n, ~n=1,2$) both subjects 
are disease-free; both life lines go to the $(1,1,0)$ direction. The first subject
gets the disease at $E$, and henceforth the life line is parallel to $(1,1,1)$ until
death at $D_1$. The second subject remains without the disease for the whole life, 
which ends at $D_2$.

\bigskip

In order to measure the frequency of events in a population, such as onset of a 
chronic disease, the person-years methods
records the number of people who are affected and the time elapsed before the event
occurs. The person-years incidence rate $\lambda$ is estimated by
\begin{equation}\label{lambda_1}
\lambda = \frac{e}{m},
\end{equation}
where $e$ is is the number of events and $m$ is the number of person-years at risk 
\cite[p. 250ff]{Woo05}. Calendar time and age often are important determinants for 
occurrence of events and have to be taken into account. This usually is achieved by
dividing the subjects' time spent in the study into calendar time and age groups. 
Let $e_{ij}$ be the
number of events taking place while subjects are in time and age group $(i,j).$
Furthermore, let $m_{ij}$ be the total time at risk spent in this group, then 
Equation \eqref{lambda_1} becomes 
\begin{equation*}
\lambda_{ij} = \tfrac{e_{ij}}{m_{ij}}.
\end{equation*}

In the planar Lexis diagram it is clear, how the time at risk $m_{ij}$ can be obtained.
Let the time and age group $(i, j)$ be defined by Cartesian product
$S_{ij} := [t_{i-1}, t_{i}) \times [a_{j-1}, a_{j})$. Each subject whose life line 
intersects with the rectangle $S_{ij}$ contributes by its time at risk in $S_{ij}$. 
To be precise, 
$m_{ij}$ is the sum of all the subjects' times at risk spent in $S_{ij}$:
\begin{equation}\label{m_1}
m_{ij} = \sum_{n=1}^N \ell^{(n)}_{ij},
\end{equation}
where $\ell^{(n)}_{ij}$ is the time at risk of subject $n, ~n=1, \dots, N,$ in the
rectangle $S_{ij}.$

Again, these ideas can be generalized to three-dimensional case: Then, $e_{ijk}$ is the
number of events taking place in the rectangular hexahedron
$$S_{ijk} := [t_{i-1}, t_{i}) \times [a_{j-1}, a_{j}) \times [d_{k-1}, d_{k}).$$ 
For the times at risk $m_{ijk}$ it holds
\begin{equation}\label{m_2}
m_{ijk} = \sum_{n=1}^N \ell^{(n)}_{ijk},
\end{equation}
where $\ell^{(n)}_{ijk}$ is the time at risk of subject $n$ in volume element $S_{ijk}.$

Given a certain study population of size $N$, the question arises how the 
subjects' contributions $\ell^{(n)}_{ijk}$
to the overall time at risk $m_{ijk}$ spent in $S_{ijk}$ can be calculated. Since
$N$ may be large (up to several thousand), some attention should be paid to 
computation time. 

The solution is straightforward by noting that the question is 
very similar to
the problem of following a radiological path through a voxel grid in tomography or 
raytracing in computer graphics. For both fields, tomography and raytracing, 
there is an ongoing research effort to efficiently discretize continuous lines 
(radiological paths or rays of light). This article has been inspired by
the seminal work of Siddon, \cite{Sid85}. 

\section{Intersecting life lines with voxels in the Lexis diagram}
Since the algorithm presented in this section is motivated from the field of
computer tomography, some of the terminology is useful. Typically one of the
sets $S_{ijk}$ resulting from a partition of
a rectangular hexahedron (right cuboid) into congruent volume elements, is called
a \emph{voxel}. The six faces of each voxel are
subsets of two adjacent planes parallel either to the $t$-$a$-plane, $a$-$d$-plane or 
$t$-$d$-plane.
Hence, the voxel space comes along with
a set of equidistant, parallel planes which are perpendicular to the abscissa, 
ordinate or applicate and which are defined by the union
of all voxel faces. These planes play a crucial role in the algorithm.

In this article all voxels $S_{ijk}$ are considered to be 
cubical, with all edges having the length $t_r, ~t_r > 0:$
\begin{equation}\label{voxel}
S_{ijk} :=  \left [ t_r \cdot \left (i-1 \right ), ~t_r \cdot i \right ) \times 
            \left [ t_r \cdot \left (j-1 \right ), ~t_r \cdot j \right ) \times 
            \left [ t_r \cdot \left (k-1 \right ), ~t_r \cdot k \right ).
\end{equation}

These voxels form a grid where the life lines of all subjects in the study are
sorted into. As a consequence of cubical voxels, the temporal 
resolution with respect to calendar time, age and duration is the same. However, 
generalization to partitions usings rectangular voxels with height, length and 
depth being different is easily possible. 

The main idea for calculating the $\ell^{(n)}_{ijk}$ in the life line ${\cal L}_n$ of 
subject $n$ starting at entry point $B_n := (t^{(n)}_0, a^{(n)}_0, d^{(n)}_0)$, 
ending at exit point $D_n := (t^{(n)}_1, a^{(n)}_1, d^{(n)}_1)$, is the parameterization
in the form
\begin{equation*}
   {\cal L}_n: ~B_n + \alpha \cdot (D_n - B_n), ~\alpha \in [0, 1].
\end{equation*}
Note, that $t^{(n)}_1 - t^{(n)}_0 = a^{(n)}_1 - a^{(n)}_0 = d^{(n)}_1 - d^{(n)}_0 =: 
\Delta t^{(n)}.$
Using this parameterization, all parameters $\alpha^{(n)} \in [0, 1]$ are calculated where
an intersection with a voxel face takes place. Since the voxels are arranged in
a regular grid, intersecting one of the voxel faces is equivalent with intersecting
one of the $t$-$a$-, $a$-$d$- or $t$-$d$-planes formed by the union of all voxel
faces mentioned above. Hence, we calculate the intersections with these planes.

Let us start with the $a$-$d$-planes (perpendicular to the $t$-axis): all those
$\alpha^{(n)}_t$ where an intersection with an $a$-$d$-plane occurs are given by
\begin{equation*}
   \alpha^{(n)}_t(u) = \frac{u \cdot t_r - (t^{(n)}_0 ~\% ~t_r)}{\Delta t^{(n)}}, ~u = 1, \dots, U^{(n)},
\end{equation*}
where $\%$ is the modulo-operator and $U^{(n)}$ denotes 
the number of intersected $a$-$d$-planes:
\begin{equation*}
U^{(n)} =  \left \lfloor \nicefrac{t^{(n)}_1}{t_r} \right \rfloor 
         - \left \lfloor \nicefrac{t^{(n)}_0}{t_r} \right \rfloor.
\end{equation*}

Similar formulas hold for those $\alpha^{(n)}_a(v), ~v=1, \dots, V^{(n)},$ and 
$\alpha^{(n)}_d(w), ~w=1, \dots, W^{(n)},$ where ${\cal L}_n$ intersects
the $t$-$d$- or $t$-$a$-planes, respectively. Now define the set
\begin{eqnarray}
A_n :=  & & \{ \alpha^{(n)}_t(u)~\vert ~u = 1, \dots, ~U^{(n)} \} \nonumber \\
        &\cup& \{ \alpha^{(n)}_a(v)~\vert ~v = 1, \dots, ~V^{(n)} \} \label{alphaSet} \\
        &\cup& \{ \alpha^{(n)}_d(w)~\vert ~w = 1, \dots, W^{(n)} \} \nonumber,
\end{eqnarray}
which contains those $\alpha^{(n)} \in [0, 1]$ where an intersection 
occurs. Note 
that the three sets on the right-hand side of Equation \eqref{alphaSet} are not 
necessarily disjoint. 
Multiple values occur if an intersection happens to be on an edge or vertex of a 
voxel. Let $A^\star_n := A_n \cup \{0, ~1\}$ be ordered ascendingly
$A^\star_n = \{\alpha^{(n)}(p) ~\vert ~p = 1, \dots, P^{(n)} \}$ with 
$0 = \alpha^{(n)}(1) < \dots < \alpha^{(n)}(P^{(n)}) = 1.$ For calculating the $\ell^{(n)}$
and the associated voxel indices $i, j, k,$ we have following algorithm:
\begin{enumerate}
\item For each subject $n, ~n=1, \dots, N,$ calculate the set $A^\star_n$ as above and sort the elements 
$\alpha^{(n)}(p), ~p = 1, \dots, P^{(n)},$ in ascending order.
\item For $p= 1, \dots, P^{(n)}$ set
\begin{equation}\label{indices}
(i_p, j_p, k_p) := \left \lfloor \frac{B_n + \alpha^{(n)}(p) \cdot (D_n - B_n)}{t_r} \right \rfloor,
\end{equation}
where the division is taken componentwise.
\item Then, calculate
\begin{equation}\label{ell}
\ell^{(n)}_{i_p j_p k_p} = \left ( \alpha^{(n)}(p+1) - 
\alpha^{(n)}(p) \right ) \cdot \Delta t^{(n)}, ~p = 1, \dots, P^{(n)} - 1.
\end{equation}
\end{enumerate}

For each voxel $(i, j, k)$ summing up all the times $\ell^{(n)}_{ijk}, ~n=1, \dots, N,$ 
by Equation \eqref{m_2} yields
the time at risk $m_{ijk}$ in period-, age- and 
duration class $(i, j, k),$  which can be used in the person-years method.

\bigskip

The idea of calculating the intersection points with the voxel faces goes back
to Robert L. Siddon. The algorithm proposed by Siddon has been developed for 
raytracing in tomography, where
several millions of paths have to be computed to form a radiological image. While
the implementation provided with this article is not tuned for efficiency, 
remarkable speeding up is possible \cite{Chr98}. The execution of
500 runs of an R (The R Foundation for Statistical Computing) 
implementation of the algorithm with the data of 200 patients 
(equivalent to a total of $10^5$ patients) 
on a 2.6 GHz personal computer takes 92 seconds.
The (simulated) patient data set is described in more detail in the next section.

\section{Examples}

An implemtentation of the method in this article has been tested with simulated
patient data. A study population of 200 subjects with entry age 55 to 85 born 
between 0 and 15 (TUs) suffering from a chronic disease for 3 to 15 TUs at 
is the time of entry is assumed to have the mortality rate
$$m(a, d) = \exp(-10 + 0.1 \cdot a) \cdot (1 + 0.1\cdot d).$$
Exit from the study is assumed to be only due to death (no censoring). The aim 
is to unfold the mortality rate from these patient data.

\newpage

\noindent For setting up the data set, following code has been used: 
\begin{verbatim}
for(patNr in 1:200){
   thisPatInAge       <- runif(1, 55, 80)
   thisPatBirth       <- runif(1,  0, 15)
   thisPatDuration    <- runif(1,  3, 15)

   F                  <- fct_F(thisPatInAge, thisPatDuration)
   thisPatDeathAge    <- round(   which.min(runif(1) > F) + thisPatInAge 
                               + (runif(1) - 0.5), 3)
   patMatrix[patNr, ] <- c(thisPatBirth, thisPatInAge, thisPatDuration,
                             thisPatDeathAge)
}
\end{verbatim}
For the simulation of the age of death, inverse transform sampling is used: \\ 
(\verb"which.min(runif(1) > F)"). Therefor the
cumulative distribution function $F(t ~\vert ~a_0, d_0)$
for someone who enters the study at age $a_0$ and having got the disease for 
$d_0$ TUs is calculated in the function \verb"fct_F":
$$F(t ~\vert ~a_0, d_0) = 1 - \exp \left ( - \int_0^t m(a_0 + \tau, d_0+\tau) \mathrm{d} \tau \right ).$$

\begin{figure}[ht]
\begin{center}
\includegraphics[width=9.6cm, keepaspectratio]{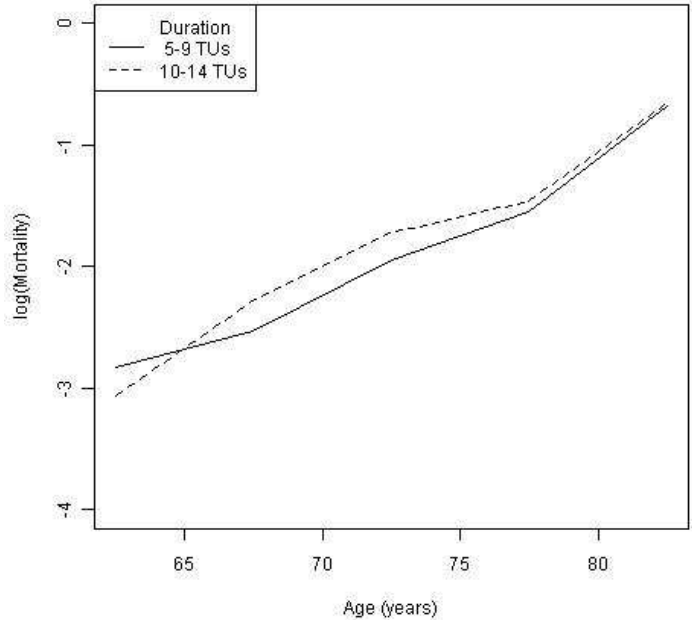}
\end{center}
\caption{Age- and duration specific mortality as estimated by the person-years 
method.}
\label{fig:Mort}
\end{figure}

If the alorithm is applied to this data set with $t_r = 5,$ 
it is possible
to estimate the mortality rate $m(a, d)$ by the person-years method.
The result is shown in Figure \ref{fig:Mort}.

\newpage

\section{Conclusion}
This article is about an extension of the person-years when period-, age- and duration-effects
occur. The method to calculate the time at risk is based on raytracing techniques used in 
tomography and provides 
a fast way to follow the individual life lines of subjects in the Lexis diagram. The
algorithm is able to treat two cases
\begin{enumerate}
\item calculate the time at risk for newly incident cases, and
\item calculate the time at risk, where the time elapsed after an event 
(e.g. duration since onset of a disease) is relevant.
\end{enumerate}
In the first case, the life lines are located in the $t$-$a$-plane, in the second the
life lines are pointing to the $(1,1,1)$ direction. The later case is important, when
the duration is a covariable. This might be the case
in late sequalae or mortality after having got a disease. Similarly, by interpreting 
the applicate axis (z-axis) as duration of exposure to a risk factor, the time at 
risk depending on period, age and duration of exposure can be calculated.

In large populations or register data ($\log_{10} N > 5$ , times at risk 
are usually estimated by a 
formula going back to Sverdrup, \cite[Sect. 3.2.]{Car06}. It is noted that using
the method described in this article, \emph{estimation} is no longer necessary, because 
given entry 
and exit times of the subjects, \emph{calculation} of times at risk is possible at
feasible computational expense.

\bibliographystyle{plain}
\bibliography{path}

\begin{thebibliography}{1}

\bibitem{Car06}
Bendix Carstensen.
\newblock {Age-Period-Cohort Models for the Lexis Diagram}.
\newblock {\em Statistics in Medicine}, 26(15):3018--3045, 2007.

\bibitem{Chr98}
Mark Christiaens, Bjorn~De Sutter, Koen~De Bosschere, Jan~Van Campenhout, and
  Ignace Lemahieu.
\newblock A fast, cache-aware algorithm for the calculation of radiological
  paths exploiting subword parallelism.
\newblock In {\em Journal of Systems Architecture, Special Issue on Parallel
  Image Processing}, 1998.

\bibitem{Kei90}
Niels Keiding.
\newblock Statistical interference in the lexis diagram.
\newblock {\em Philosophical Transactions of the Royal Society London A},
  332:487--509, 1990.

\bibitem{Kei91}
Niels Keiding.
\newblock Age-specific incidence and prevalence: a statistical perspective.
\newblock {\em Journal of the Royal Statistical Society A}, 154:371--412, 1991.

\bibitem{Kei06}
Niels Keiding.
\newblock Event history analysis and the cross-section.
\newblock {\em Statistics in Medicine}, 25(14):2343--2364, 2006.

\bibitem{Sid85}
Robert~L. Siddon.
\newblock {Fast Calculation of the Exact Radiological Path for a
  Three-Dimensional CT Array}.
\newblock {\em Medical Physics}, 12(2):252--255, 1985.

\bibitem{Woo05}
Mark Woodward.
\newblock {\em Epidemiology: Study Design and Data Analysis}.
\newblock Texts in statistical science. Chapman \& Hall/CRC, 2005.

\end{thebibliography}

\end{document}